\documentclass[aps,prb,reprint,superscriptaddress]{revtex4-1}

\usepackage{amsmath}
\usepackage{bm}
\usepackage{tabularx}
\usepackage{graphicx}
\begin{document}
\newcommand{\tn}{$T_{\rm N}$}
\newcommand{\uas}{UAu$_2$Si$_2$}
\newcommand{\mub}{${\mu}_{B}$}

\title{Neutron Diffraction Study on Single-crystalline UAu${_2}$Si$_2$}

\author{Chihiro Tabata$^1$\thanks{ctabata{\_}@post.kek.jp}, Milan Klicpera$^{2,3}$, Bachir Ouladdiaf$^2$, Hiraku Saito$^4$, Michal Vali{\v{s}}ka$^3$, Kl{\'a}ra Uhl{\'i}{\v{r}}ov{\'a}$^3$, Naoyuki Miura$^4$, Vladim{\'i}r Sechovsk{\'y}$^3$, and Hiroshi Amitsuka$^4$}

\affiliation{Institute of Materials Structure Science, High Energy Accelerator Organization, Tsukuba 305-0801, Japan \\
$^2$Institut Laue-Langevin, 71 Avenue des Martyrs, CS 20156, 38042 Grenoble Cedex 9, France \\
$^3$Charles University, Faculty of Mathematics and Physics, Department of Condensed Matter Physics, Ke Karlovu 5, 121 16 Prague 2, Czech Republic \\
$^4$Graduate School of Science, Hokkaido University, Sapporo 060-0810, Japan} 

\date{\today}

\begin{abstract}
Magnetic structure of tetragonal \uas \ was investigated by single-crystal neutron diffraction experiments. 
Below \tn = 20 K it orders antiferromagnetically with a propagation vector of $\bm{k} = (2/3, 0, 0)$ and magnetic moments of uranium ions pointing along the tetragonal $c$-axis. 
Weak signs of the presence of a ferromagnetic component of magnetic moment were traced out.
Taking into account a group theory calculation and experimental results of magnetization and $^{29}$Si-NMR, the magnetic structure is determined to be a squared-up antiferromagnetic structure, with a stacking sequence ($+ + -$) of the ferromagnetic $ac$-plane sheets along the $a$-axis.
This result highlights similar magnetic correlations in \uas \ and isostructural URu$_2$Si$_2$. 
\end{abstract}

\maketitle

\section{Introduction}
U$T_2$Si$_2$ ($T$: transition metal) compounds have provided fruitful opportunities for systematic study of 5f-electron properties in strongly-correlated electronic systems, which have been attracting much interest with a variety of phenomena such as heavy-fermion states, superconductivity, magnetic ordering, and hidden order. 
Up to date, thirteen stable U$T_2$Si$_2$ compounds with different $T$ elements have been confirmed to exist, and their 5f-electronic ground states have been identified through a spectrum of experiments, except for two systems: URu$_2$Si$_2$ and \uas. 
The former is well known to exhibit the hidden order transition at 17.5 K \cite{Palstra85, Maple86, Schlabitz86, Mydosh_Rev14}, being studied intensively for several decades. 
In contrast, the latter has only five reports which were made between the 80s and 90s. Although a ferromagnetic ground state below about 20 K was suggested in those reports\cite{Palstra86, Saran88, Rebelsky91, Torikachvili92, Lin97}, a major part of the detailed magnetic properties were unknown since all the studies were done on polycrystalline samples. 

Nearly 20 years after the last report \cite{Lin97}, we succeeded in growing single-crystalline samples of \uas. It crystallizes in the ThCr$_2$Si$_2$ type of tetragonal structure (space group: $I4/mmm$, $D_{4h}^{17}$), like most U$T_2$Si$_2$ compounds. 
Our detailed magnetization measurements revealed peculiar behaviors of this compound \cite{Tabata16}.
In a temperature range from room temperature to $\sim$ 50 K, the magnetic susceptibility shows the Curie-Weiss behavior for both field directions along the $a$-axis and the $c$-axis, yielding an almost isotropic effective moment.
Below $\sim$ 50 K, the susceptibility becomes highly anisotropic by an emergence of a weak ferromagnetism along the $c$-axis.
The magnetic anisotropy gets more remarkable below the transition temperature of $\sim$ 20 K by emergence of another ferromagnetic (FM) component along the $c$-axis, while the basal-plane susceptibility is suppressed.
Interestingly, an antiferromagnetic (AFM) component, which is masked by the FM component, becomes conspicuous by applying a magnetic field. It means that the ground state of \uas \ is not simply FM but intrinsically AFM. Recent $^{29}$Si-NMR experiments also provided an evidence of the AFM order \cite{Tabata17-1}.

In the present paper, we focus on the magnetic structure taking place below \tn $\sim$ 20 K in \uas. 
We report results of neutron diffraction experiments on a single crystalline sample.
Observed magnetic reflections, together with the bulk magnetization and the $^{29}$Si-NMR\cite{Tabata16, Tabata17-1}, indicate that the magnetic structure is spin-uncompensated AFM with $\bm{k} = (2/3, 0, 0)$ and magnetic moments arranged along the $c$-axis, which points out an intriguing similarity to the isostructural URu$_2$Si$_2$ \cite{Kuwahara13, Knafo16}. 

\section{Experimental Procedure}
UAu$_2$Si$_2$ single crystal was prepared by the floating zone melting method in a four-mirror optical furnace (Crystal Systems Co.) from a polycrystalline precursor. The grown crystal was characterized by the Laue X-ray diffraction, energy dispersive X-ray spectrometry (EDX), and measurements of magnetization and electrical resistivity. For the neutron diffraction experiments, a rectangular piece with dimensions of $\sim$ 1 mm $\times$ 1 mm $\times$ 2 mm was cut out of the grown rod-shaped crystal. The sample was attached to an Al holder with GE varnish.
First, we performed a Laue neutron diffraction experiment using CYCLOPS diffractometer at Institut Laue-Langevin (ILL), Grenoble.
A double octagonal array of neutron CCD detectors covers a cylindrical area of space \cite{Ouladdiaf11}.
The scans were taken at temperatures from 2 K up to 70 K, mapping both magnetically ordered and paramagnetic state.
Subsequently, the single crystal was investigated using the four-circle diffractometer D10 (ILL) employing an area detector. The sample $a$-axis was parallel to the omega-axis of the diffractometer.
The sample was cooled down to 2 K by a $^ 4$He gas-flow refrigerator. The half-lambda component was removed by a graphite filter; its intensity was reduced to 10$^{-4}$ of the main component.
At 30 K (above \tn), about 100 independent nuclear reflections were measured for a refinement of the crystal structure using incident neutron wavelength of $\lambda$ = 1.26 \AA.  
The measurements of magnetic reflections ($\sim$ 250 independent magnetic reflections) were performed using incident neutron wavelength of $\lambda$ = 2.36 \AA \ at 2 K.
Nuclear reflections were also measured at 2 K, in order to detect $\bm{k} = 0$ components of the magnetic moments.
The least-square refinements of data to model crystal and magnetic structures were carried out using the FullProf package \cite{FullProf, WinPLOTR}. The detailed D10 data can be found in Ref. \cite{Klicpera16}.

\section{Results}
\begin{figure}[h]
\includegraphics[width=8.5cm]{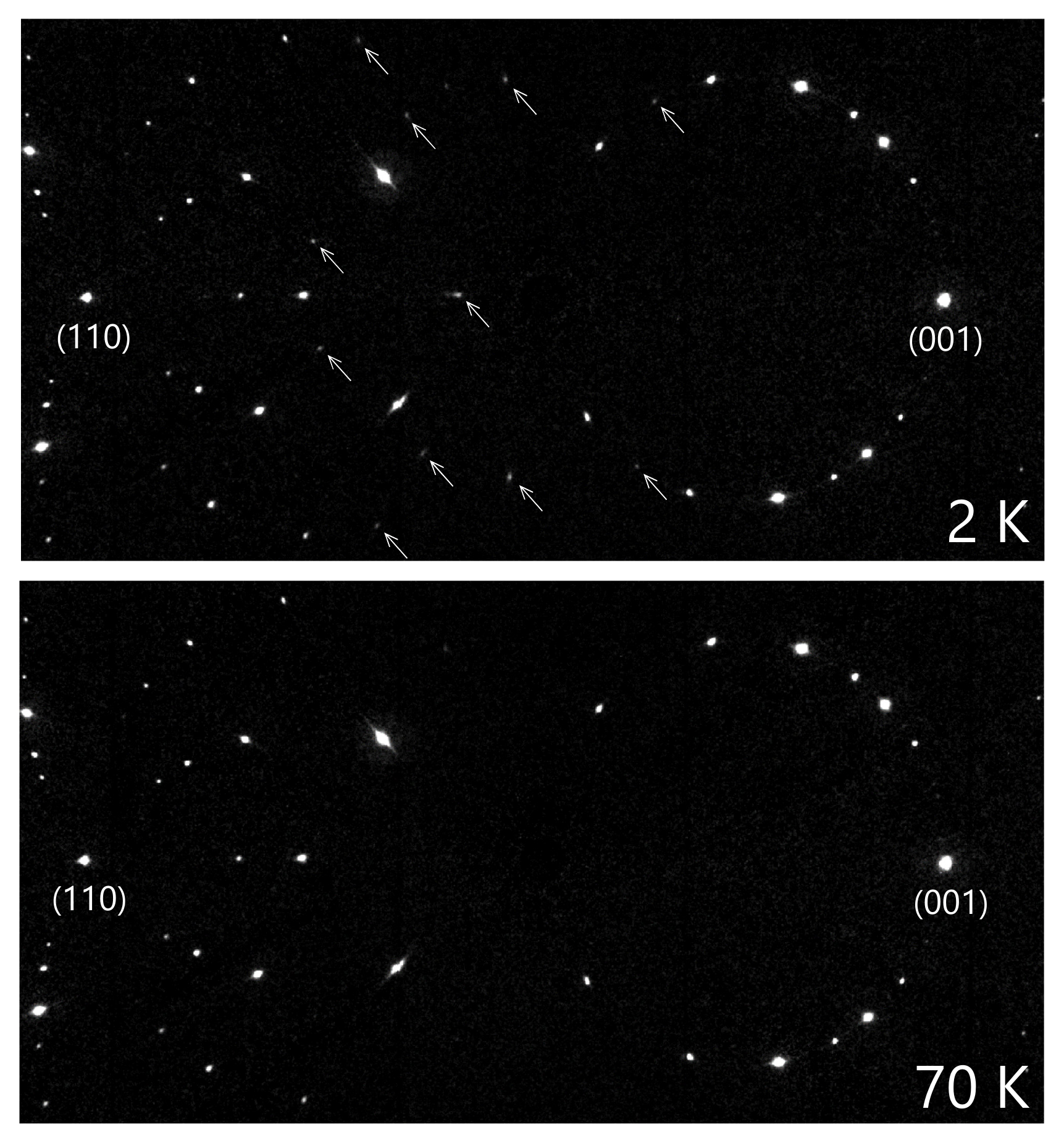}
\caption{Neutron Laue diffraction patterns of UAu$_2$Si$_2$, acquired at temperatures of 2 K (below $T_{\rm N}$) and 70 K (above $T_{\rm N}$). The magnetic reflections are marked by arrows.}
\label{Laue}
\end{figure}

A comparison of Laue neutron diffraction patterns (CYCLOPS) taken in the paramagnetic state and the ordered state, i.e. at 70 K and 2 K, revealed a number of magnetic reflections outside of the Bragg positions of nuclear structure.
Thus the AFM order below \tn \ suggested by $^{29}$Si-NMR \cite{Tabata17-1} was unambiguously confirmed.
The observed reflections are well described by the propagation vector of $\bm{k}$ = (0.67, 0, 0), which is close to (2/3, 0, 0), and the vector star.
The Laue images acquired at 2 K (below \tn) and 70 K (above \tn) are shown in Fig. \ref{Laue}.
No higher-order satellite was detected.

\begin{figure}[h]
\includegraphics[width=8.5cm]{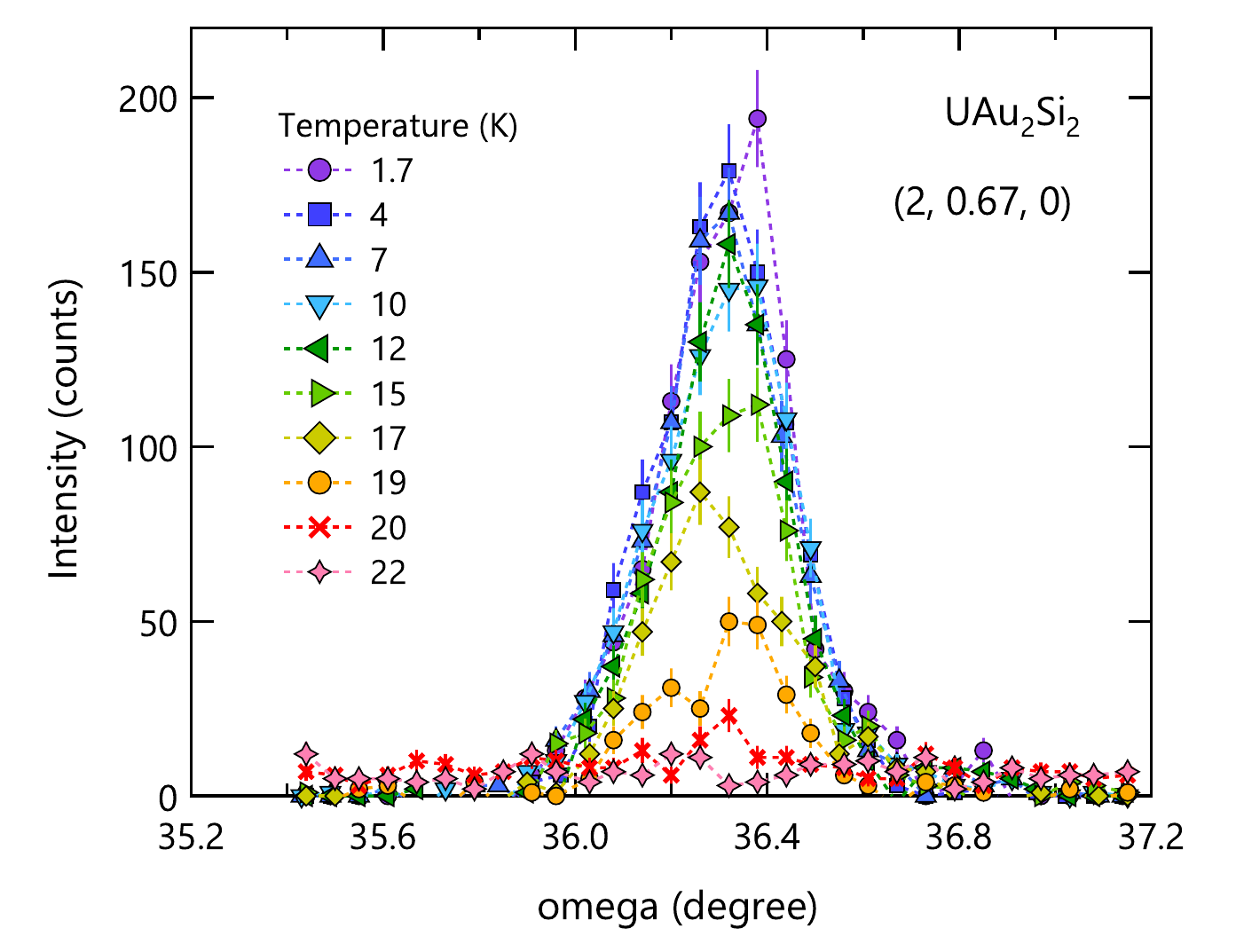}
\caption{(Color online) Temperature dependence of peak profiles of the reflection at $\bm{q} = (2, 0.67, 0)$.}
\label{profile}
\end{figure}

\begin{figure}[h]
\includegraphics[width=8.5cm]{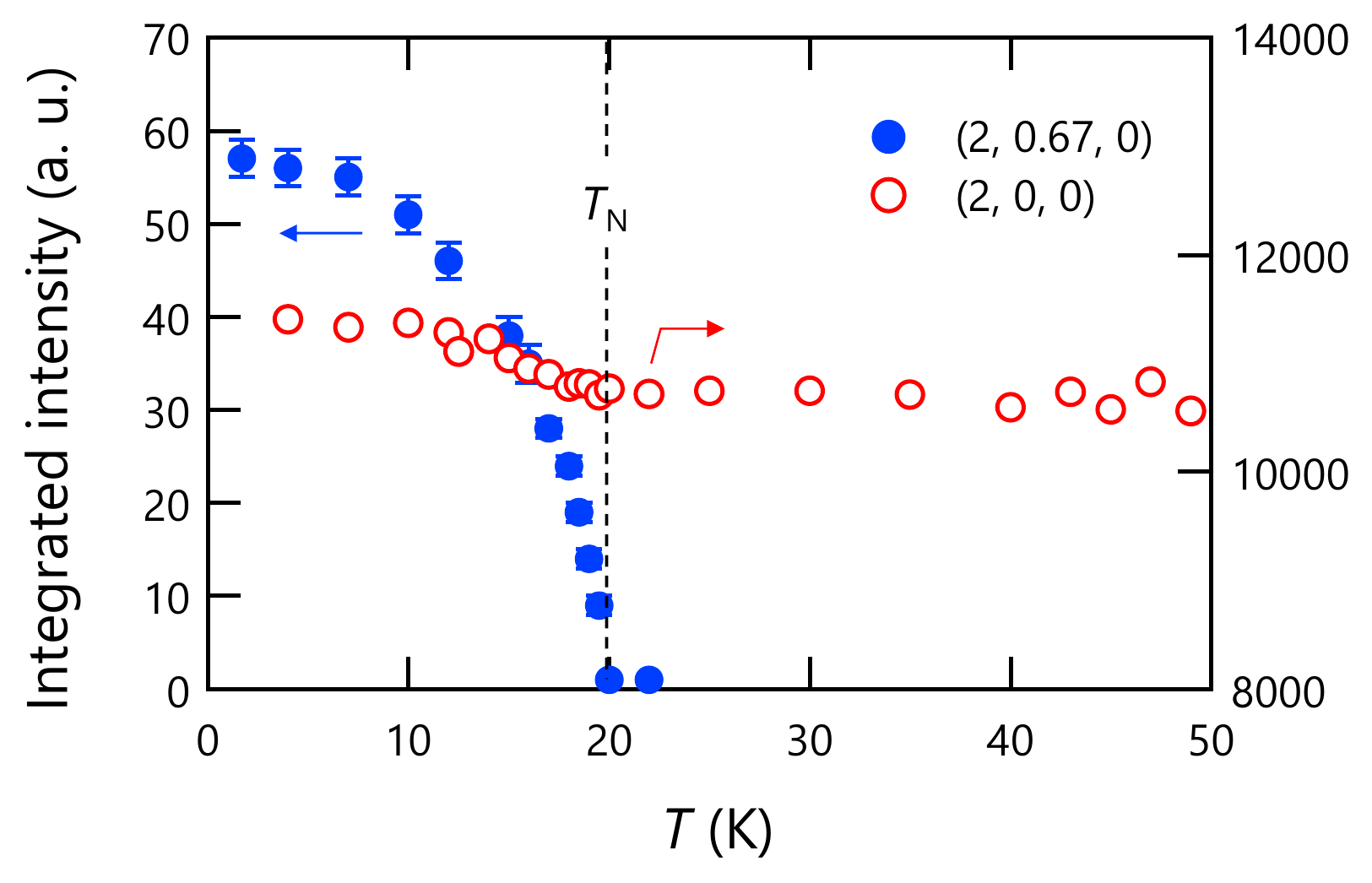}
\caption{(Color online) Temperature dependence of the integrated intensity of the peaks at (2, 0.67, 0) (closed circles) and (2, 0, 0) (open circles).}
\label{T-dep_int}
\end{figure}

Figure \ref{profile} shows peak profiles of a superlattice reflection at $\bm{q} = (2, 0.67, 0)$ measured on D10.
The intensity of the magnetic peak decreases with increasing temperature, and vanishes when the temperature reaches \tn.
The temperature dependence of its integrated intensity is shown in Fig. \ref{T-dep_int}, together with that of a fundamental nuclear reflection peak at  $\bm{q} = (2, 0, 0)$.
The intensity of the (2, 0.67, 0) reflection increases below \tn, and almost saturates below 10 K.
Simultaneously it is notable that the intensity of the (2, 0, 0) reflection also increases below \tn, appearing to be due to the FM component, and saturates below 10 K.
However, the increase in intensity is nearly 100 times larger than that accounted for by a magnetic origin assuming a magnetic moment of $\sim$1 \mub \ per U ion. 
Then the first scenario to be considered is a structural transition simultaneously occurring at \tn, but this is unlikely or it is very subtle even if it does occur; we have never observed any signs of peak broadenings or changes of peak profiles in either the present neutron scattering experiments or our previous powder X-ray diffraction experiments \cite{Tabata16}.
Our recent measurement of thermal expansion revealed the temperature variations of the lattice parameters below \tn \cite{Michal17}.
However, these variations are of the order of 10$^{-4}$, i.e. they are too small to be detected by those diffraction experiments.
The other scenario may be a change of the extinction effect by a magnetic domain formation.
In general, the extinction effect weakens peak intensity, but this effect can be restrained in a crystal which forms magnetic domains. 
We found that an increase of the peak intensities is remarkable for the stronger peaks in the present experiment, which is consistent with a general fact that the extinction effect is more significant for stronger reflections.

 \renewcommand{\arraystretch}{1.3}
\begin{table*}[t]
\caption{Structural parameters of UAu$_2$Si$_2$ deduced from the D10 data measured at 30 K. Values of $U_{\rm eq}$ are from the X-ray diffraction data of isostructural compound URu$_2$Si$_2$ \cite{Tabata_URS} used for the final fit as described in text. }
\label{table_lattice}
\begin{center}
\vspace{0.2cm}
\begin{tabular*}{17.5cm}{p{3.5cm}p{1cm}p{3cm}p{2cm}p{1.3cm}p{1.3cm}p{1.5cm}l}
\hline
Lattice parameter (\AA)& Site & Wyckoff position & symmetry & \multicolumn{3}{c}{position}& $U_{\rm eq} \times 10^3$ (\AA) \cite{Tabata_URS}\\
                 &         &                          &                &   {\it x} & {\it y} & {\it z}     &                         \\
\hline
$a = 4.20(5)$   & U  & $2a$ & $4/mmm$             & 0    & 0 & 0              & 1.42(4) \\
$c = 10.26(5)$ & Au & $4d$ & $\overline{4}m2$ & 1/2 & 0 & 1/4           & 1.70(4) \\
                   &  Si & $4e$ & $4mm$                 & 0    & 0 &  0.389(1) & 2.5(2) \\
\hline
\end{tabular*}
\end{center}
\end{table*}

The ThCr$_2$Si$_2$ type crystal structure of \uas \ was refined by analyzing integrated intensity data collected at 30 K. The refined structural parameters were the $z-$coordinate of the Si atoms and the atomic thermal displacement parameters. 
The $z$-coordinate of Si was refined to be 0.389(1) well in an agreement with our previous study \cite{Tabata16}.
On the other hand, the atomic thermal displacement parameters were unable to be refined to appropriate values, because of the limited number of observed reflections particularly in the high angle region in the present neutron diffraction experiments. 
The final refinement of our results was done using fixed values for the thermal displacement parameters as determined for the isostructural compound URu$_2$Si$_2$ in our previous high-energy synchrotron X-ray diffraction experiments at 20 K \cite{Tabata_URS}. 
The used parameters improved the agreement between the data and model only slightly. The obtained structural parameters are listed in Table \ref{table_lattice}.
Three extinction parameters were applied in FullProf codes for correction of the extinction effect which causes significant deviation of the calculated structure factors of several strong reflections from observed ones.

Based on the refined crystal structure, the magnetic structure was refined using data collected at 2 K. 
Firstly we present the results of refinement computing diffraction intensities with integer $hkl$ Miller indices, i.e. the intensities originating from both the nuclear structure and magnetic structure with $\bm{k}=0$. 
In \uas, a FM component of $\mu_{\rm FM}\sim$ 0.3 \mub/U along the $c$-axis was confirmed by our previous magnetization measurements \cite{Tabata16}. 
Hence all of those diffraction intensities contain magnetic contribution, which means that there is no pure nuclear reflection at least below \tn.
However, the FM component cannot be determined in the present neutron experiment for its much smaller contributions to the neutron diffraction intensities than that of the nuclear origin (less than 10$^{-3}$ of the strongest reflection intensity); it is nearly negligible within the accuracy of the present experiment. 
Nevertheless, we observed that the agreement between data and fit decreased when taking a fixed value of $\mu_{\rm FM}$ exceeding 0.5 \mub \ into the model. 
Thus the present neutron diffraction experiment suggests $\mu_{\rm FM}$ lower than 0.5 \mub, which is consistent with 0.3 \mub \ expected from the magnetization measurement \cite{Tabata16}. 
The typical best-fit reliability factors when assuming $0 \ \mu_{\rm B}$/U $< \mu_{\rm FM} < 0.5 \ \mu_{\rm B}$/U are: $R_{F2} = 4.5 \%, R_{F2w} = 4.5 \%, R_{F} = 2.9 \%$.

Secondly, we performed the refinements of the AFM component by computing magnetic reflection intensities whose indices are non-integer; in the present case, they are reflections with $\bm{q} = \bm{Q} \pm \bm{k}_1$ or $\bm{q} = \bm{Q} \pm \bm{k}_2$, where $\bm{Q}$ is the reciprocal lattice vector and $\bm{k}_1$ and  $\bm{k}_2$ are the propagation vectors of the magnetic structure, expressed as $\bm{k}_1 = (0.67, 0, 0)$ and $\bm{k}_2 = (0, 0.67, 0)$. 
Before the refinements, we found a highly unequal distribution of the $\bm{k}_{1}$-domain and the $\bm{k}_{2}$-domain.
The estimated ratio of the volume of the  $\bm{k}_{1}$-domain and  $\bm{k}_{2}$-domain is approximately 1 : 1.8.
This inequality eliminates the possibility of a double-$\bm{k}$ structure with a single magnetic domain.
The reason for the inequality is not clear at this point, but we conjecture that it might be related to a crystallographic disorder of this compound. In order to clarify this, careful investigation of sample dependence and the crystal structure is necessary in future.

\begin{figure}[h]
\includegraphics[width=8cm]{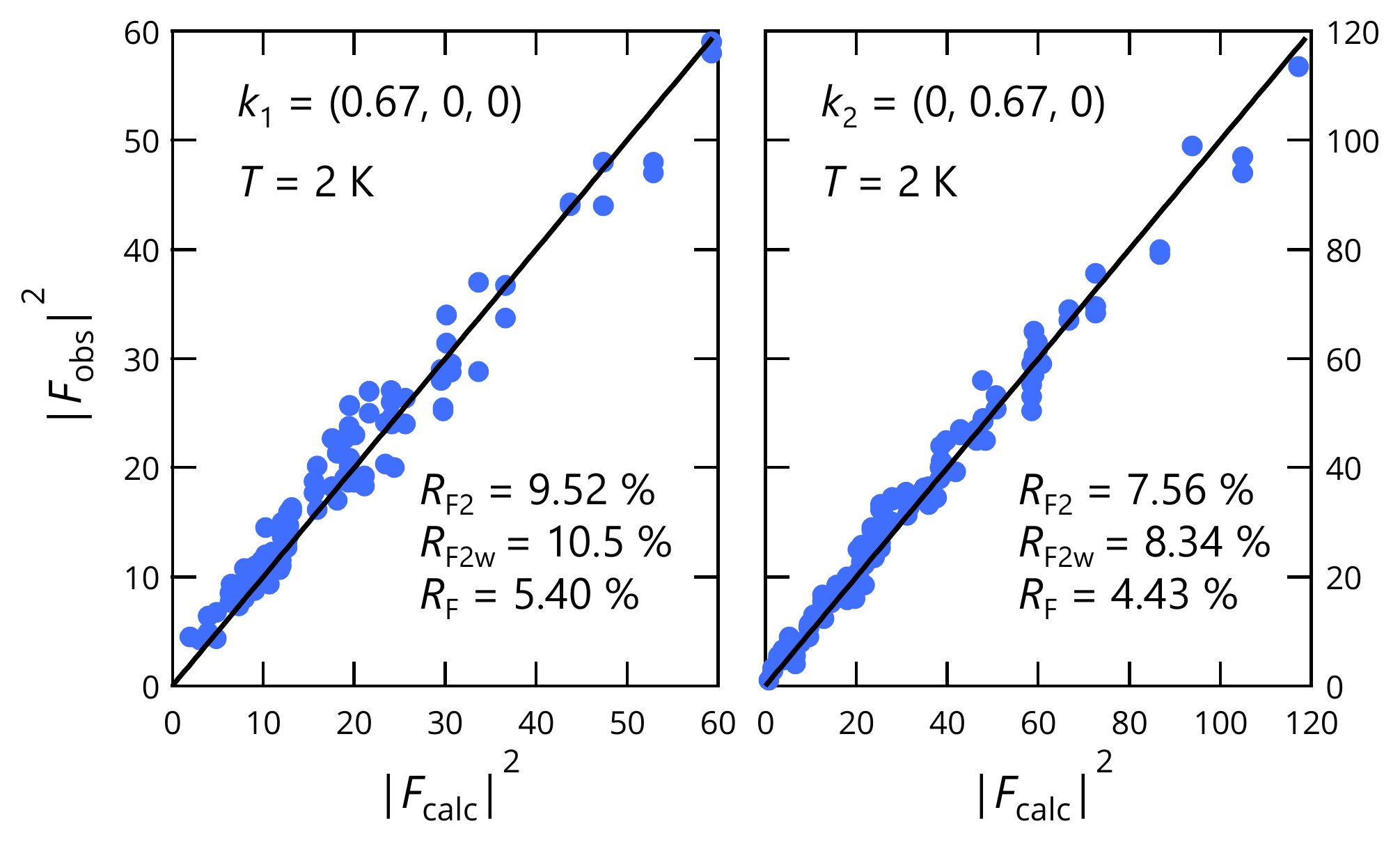}
\caption{The $F_{\rm o}$-$F_{\rm c}$ plots of magnetic reflections observed at 2 K for each magnetic domain of UAu$_2$Si$_2$. The solid lines are guides to eye, representing proportional relations of $y = x$.}
\label{FoFc}
\end{figure}

Before describing the magnetic structure model for the refinement, we summarize the results obtained from the present neutron diffraction experiment: (i) The propagation vector is $\bm{k} = (2/3, 0, 0)$ meaning that the magnetic unit cell is tripled along the $a$ axis compared to the primitive crystal lattice.  (ii) The absence of higher-$\bm{q}$ reflections suggests the sinusoidal modulation of magnetic moments. The observed integrated intensities are best refined with a sinusoidal modulation magnetic arrangement with moments aligned along the $c$-axis.
In this magnetic structure, the magnetic moment of the $i$-th U-ion $\mu_{\rm ord}(\bm{r}_i)$ is expressed as $\mu_{\rm ord}(\bm{r_i}) = M{\sin}(2\pi\bm{k} \cdot \bm{r}_i + \phi)$, where $M$ is the modulation amplitude, $\bm{r}_i$ is the position of the $i$-th U-ion, and $\phi$ is the associated magnetic phase. $M$ was refined to be $M = 1.2 \pm 0.05$ \mub/U.
The magnetic structure refinements were performed for each $\bm{k}_1$ and $\bm{k}_2$ domain, with fixed scale factors determined in accordance with the volume ratio of 1 : 1.8.
Figure \ref{FoFc} displays the $F_{\rm o}$-$F_{\rm c}$ plot, showing squared calculated structure factors versus those observed. 
The linear dependence indicates a good agreement between the measurements and the model calculation.

\begin{figure}[h]
\begin{center}
\includegraphics[width=5cm]{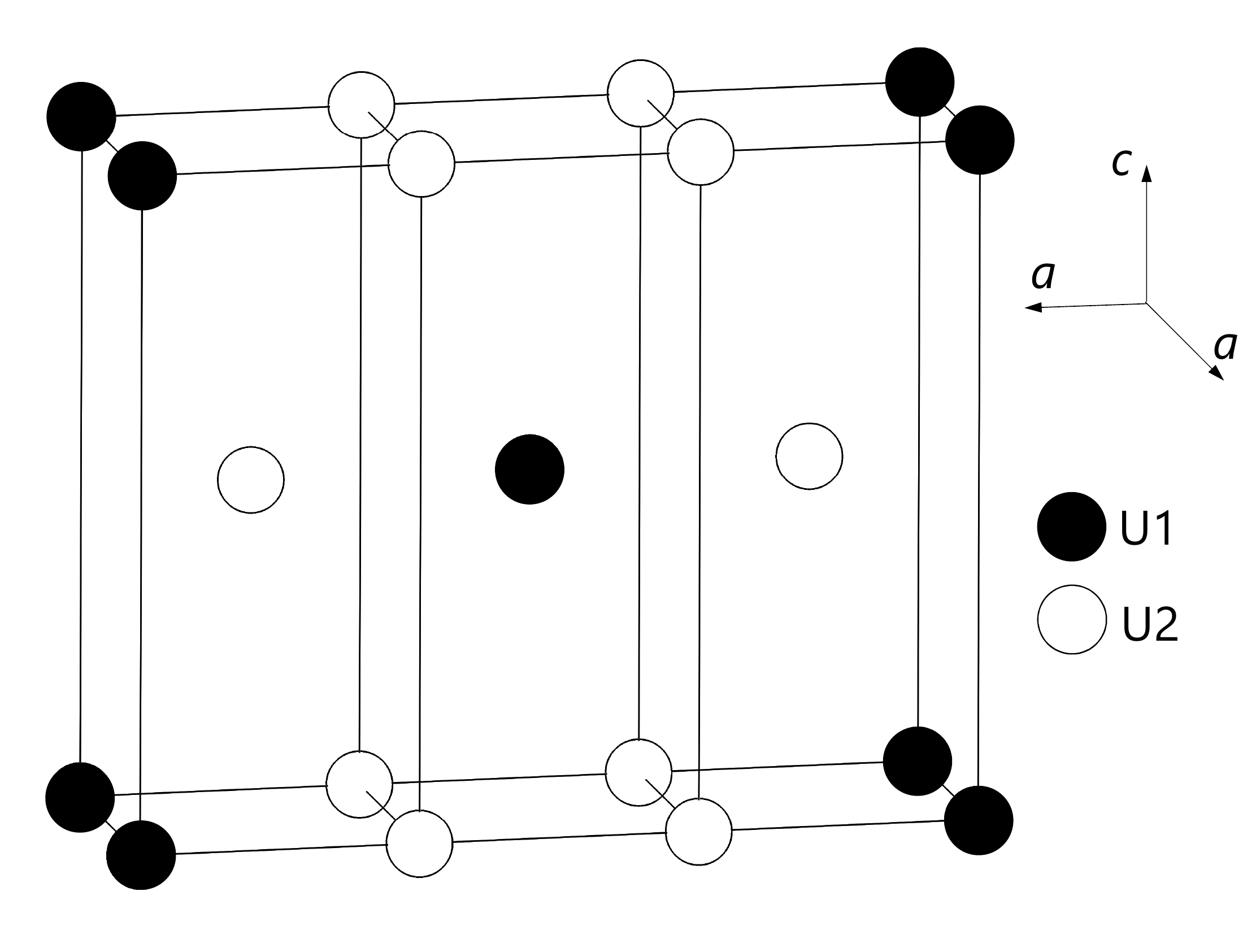}
\caption{Schematic illustration of the magnetic unit cell of the maximal magnetic subgroups for paramagnetic space group of $I4/mmm$ and a propagation vector of $\bm{k} = (2/3, 0, 0)$. Black balls and white balls specifies inequivalent magnetic sites, U1 and U2, respectively.}
\label{unit_cell}
\end{center}
\end{figure}

\begin{figure}[h]
\includegraphics[width=8.5cm]{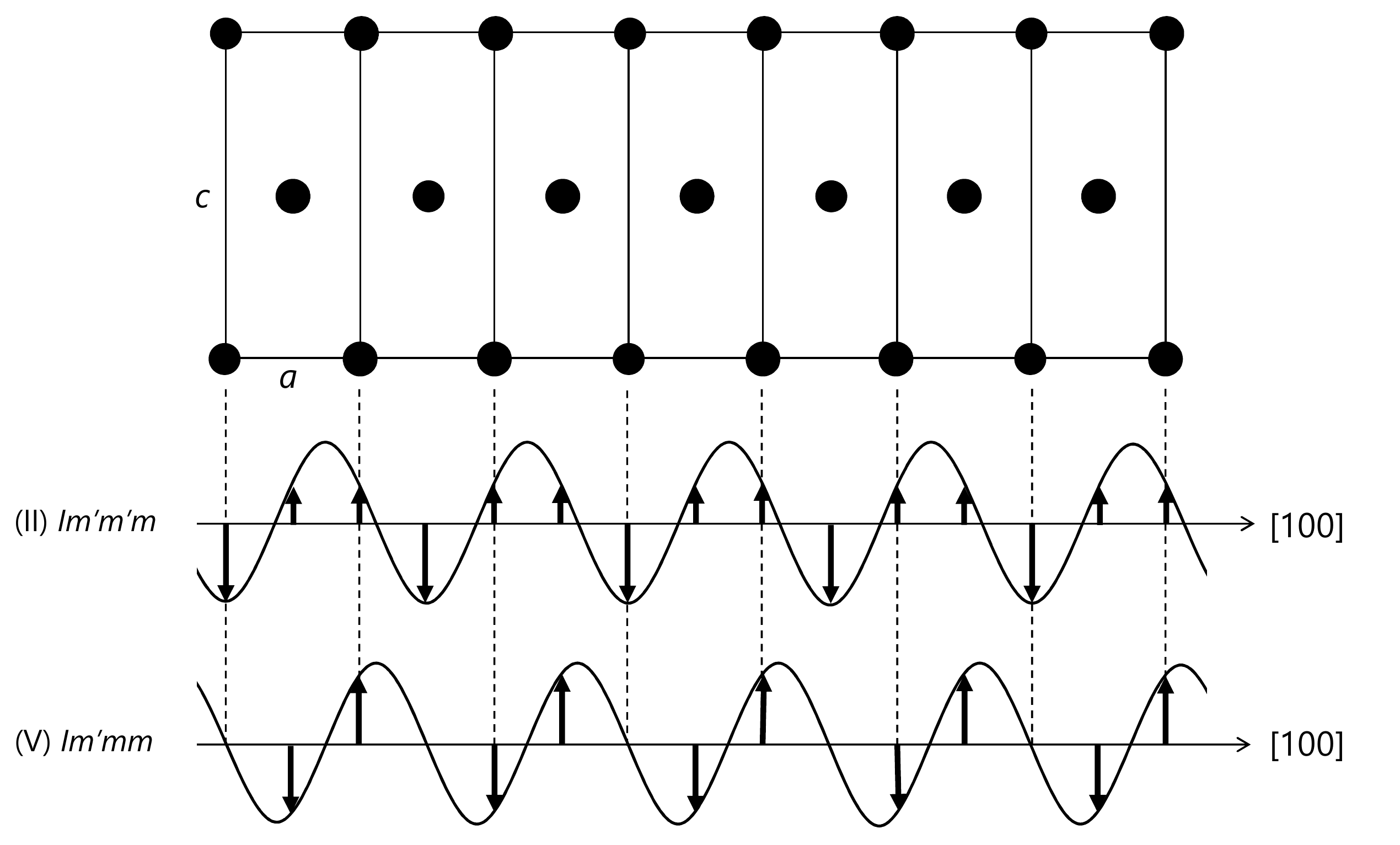}
\caption{Schematic illustration of the two patterns of spin configurations of AFM components which explains the neutron diffraction data measured on UAu$_2$Si$_2$.}
\label{structures_wave}
\end{figure}

In contrast to the amplitude $M$, the phase $\phi$ cannot be determined from experimental data alone, because diffraction data essentially lack the list of information of the phase.
Hence we narrowed down candidate magnetic structures from a viewpoint of the group theory, considering a continuous second-order transition.
A group theory analysis taking paramagnetic $I4/mmm$ space group with uranium atoms at the 2$a$ site (see Table \ref{table_lattice}) and a (2/3, 0, 0) propagation vector leads to six maximum magnetic subgroups, which allow non-zero magnetic moments.
The magnetic unit cell of these structures is tripled along the $a$ direction compared with the crystallographic unit cell, and contains two nonequivalent U sites, which we refer to U1 and U2 here, as listed in Table \ref{table_positions}.
The corners and the body center position of the magnetic unit cell belong to U1 site, and the rest of uranium positions are of the U2 site (see Fig. \ref{unit_cell}).
Table \ref{space_group} lists the magnetic space groups with candidate magnetic structures as calculated using MaxMagn program \cite{MaxMagn}.
The best-fit solutions were obtained on the two sine-wave modulated structures which allow only the $z$-component of magnetic moments, the structures of (II) $Im'm'm$ and (V) $Im'mm$ in Table \ref{space_group}, as illustrated in Fig. \ref{structures_wave}. The magnetic phase $\phi$ has fixed values in these models, which are $-{\pi}/2$ and $-\pi$ for the structures II and V, respectively.
Only $\phi$ differs by $\pi/2$ between these structures, which means that it is impossible to distinguish between them by neutron diffraction; the fitting analyses yield essentially the same results.

\renewcommand{\arraystretch}{1.5} 
\begin{table}[h]
\caption{List of magnetic sites in the magnetic unit cell shown in Fig. \ref{unit_cell}. The Wyckoff positions are defined in the magnetic unit cell which is composed of crystallographic unit cells tripled along the $a$-axis.}
\label{table_positions}
\begin{center}
\vspace{0.2cm}
\begin{tabular}{lp{4cm}l}
\hline
Site & New Wyckoff position & Multiplicity \\
\hline
U1 & (0, 0, 0), (1/2, 1/2, 1/2) & 2 \\
U2 & (1/6, 1/2, 1/2),  (1/3, 0, 0), (2/3, 0, 0), (5/6, 1/2, 1/2) & 4 \\
\hline
\end{tabular}
\end{center}
\end{table}

\begin{table}[h]
\caption{List of maximal magnetic subgroups for the paramagnetic space group $I4/mmm$ and the propagation vector $\bm{k} = (2/3, 0, 0)$. The $xyz$-coordinates corresponds to the crystallographic $abc$ axes.}
\label{space_group}
\begin{center}
\begin{tabular}{p{1cm}p{1.5cm}p{2cm}p{2cm}}
\hline
Label & Magnetic subgroup & \multicolumn{2}{l}{Magnetic moment}\\
      &                      & U1                & U2 \\
\hline
I & $Im'm'm'$  & (0, 0, 0) & ($M_{2x}$, 0, 0) \\
II & $Im'm'm$  & (0, 0, $M_{1z}$) & (0, 0, $M_{2z}$) \\
III & $Im'm'm$  & (0, $M_{1y}$, 0) & (0, $M_{2y}$, 0) \\
IV & $Im'm'm$ & ($M_{1x}$, 0, 0) & ($M_{2x}$, 0, 0) \\
V & $Im'mm$ & (0, 0, 0) & (0, 0, $M_{2z}$) \\
VI & $Im'mm$ & (0, 0, 0) & (0, $M_{2y}$, 0) \\
\hline
\end{tabular}
\end{center}
\end{table}

\begin{figure}[h]
\begin{center}
\includegraphics[width=8.2cm]{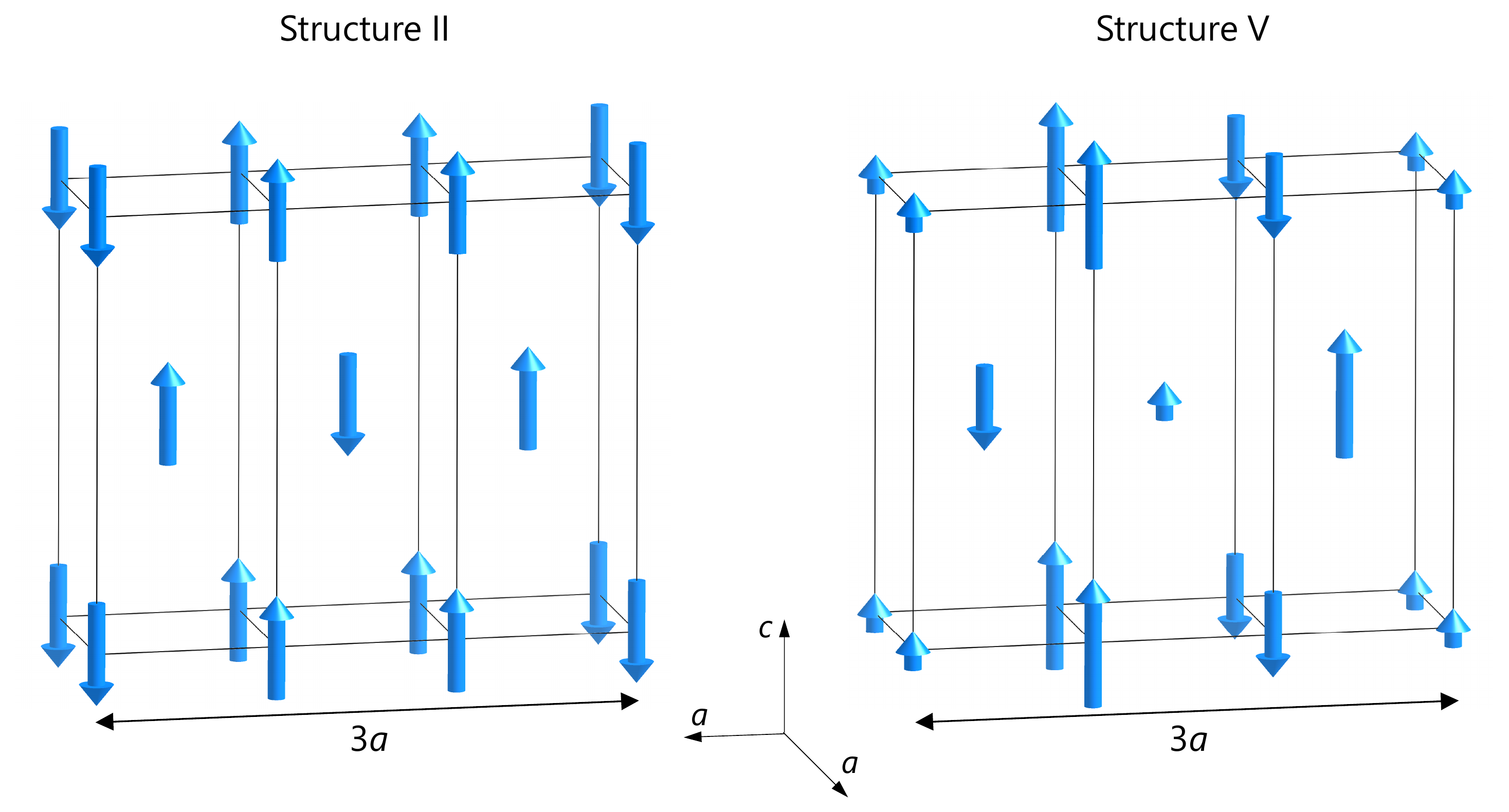}
\caption{Two candidates of the magnetic structure of UAu$_2$Si$_2$ drawn with VESTA\cite{Momma11}. They are obtained by summing up the AFM components refined by the neutron diffraction data and the FM component estimated by the magnetization data.}
\label{structures_result}
\end{center}
\end{figure}

As described above, the FM component cannot be determined specifically in the present neutron diffraction study, suggesting 0 \mub/U $<$ $\mu_{\rm FM}$ $<$ 0.5 \mub/U.
Nevertheless, the bulk magnetization indicates $\mu_{\rm FM}\sim$ 0.3 \mub/U.
Here we can choose a solution consistent with both the magnetization and the neutron diffraction results, by combining the $\mu_{\rm FM}$ of 0.3 \mub \ to the AFM component as follows.
On the structure II,
\begin{align}
&\mu_{\rm U1} \simeq |-1.2 + 0.3| = 0.9  \ (\mu_{\rm B}/{\rm U}) \nonumber \\
&\mu_{\rm U2} \simeq |0.6 + 0.3| = 0.9  \ (\mu_{\rm B}/{\rm U}). \nonumber
\end{align}
On the structure V,
\begin{align}
&\mu_{\rm U1} \simeq |0 + 0.3| = 0.3  \ (\mu_{\rm B}/{\rm U}) \nonumber \\
&\mu_{\rm U2} \simeq |\pm1.0 + 0.3| = 1.3, 0.7 \ (\mu_{\rm B}/{\rm U}), \nonumber
\end{align}
where $\mu_{\rm U1}$ and $\mu_{\rm U2}$ are ordered moments on the U1 site and the U2 site, respectively.
Figure \ref{structures_result} shows schematic views of these spin configurations.
The structure II becomes a simple squared-up ($++-$) structure, while in the structure V the modulation of magnetic moments produces three different magnitudes of magnetic moments. 

Although we cannot determine which structure is more probable by only neutron diffraction, we can rule out structure V by considering the previous results of $^{29}$Si-NMR experiments.
It is easily deduced by counting the number of magnetically-equivalent Si sites, that the NMR spectrum expected on structure V contains three distinct peaks with the intensity ratio of 1 : 1 : 1 below \tn, in a magnetic field applied parallel to the $c$-axis.
It does not agree with the observed spectrum, which contains only two peaks with the intensity ratio of 1 : 2  \cite{Tabata17-1}.
In contrast, the structure II well explains the observed spectrum.
The NMR experiments suggest a larger magnetic moment of 1.4(1) \mub/U, compared with 0.9(1) \mub/U from the neutron experiments.
This may be due to the magnetic field of 4 T applied in the NMR experiments, whereas the neutron experiment was performed in zero magnetic field.

\section{Discussion}
A magnetic structure with a propagation vector along the $a$-axis, the case of \uas \, has not been found in other U$T_2$Si$_2$ compounds in zero magnetic field.
In high magnetic field, similar magnetic structures have been observed in URu$_2$Si$_2$ and the Rh-doped system U(Ru$_{0.96}$Rh$_{0.04}$)$_2$Si$_2$.
In pure URu$_2$Si$_2$,  a spin density wave (SDW) phase with a propagation vector of (0.6, 0, 0) appears \cite{Knafo16}, when the enigmatic hidden order phase disappears in a magnetic field around 35 T applied along the $c$-axis \cite{Jaime02, Kim03}.
Interestingly, the field-induced SDW is changed into a squared-up AFM order with a propagation vector of $\bm{k} = (2/3, 0, 0)$ by 4-percent Rh doping \cite{Kuwahara13}. 
 
A variety of magnetic structures in U$T_2$Si$_2$ systems, particularly for UNi$_2$Si$_2$ and UPd$_2$Si$_2$, have been studied theoretically.
Most U$T_2$Si$_2$ compounds (with $T$ = Co, Ni, Rh, Pd, Ir, Pt) order in magnetic structures where ferromagnetic layers stack along the $c$-axis \cite{Kuznietz96,Lin91,Bak81,Shemirani93, Honma98,Verniere96,Bak85}.
Those magnetic orders have been treated based on the axial next-nearest neighbor Ising (ANNNI) model \cite{Chelmicki85, Bak81, Shemirani93, Verniere96, Bak85}.
 For example, the magnetic-field-temperature phase diagram of UPd$_2$Si$_2$ was successfully explained by the ANNNI model with a Landau-type analysis \cite{Plumer94, Honma98}.
This kind of analysis also worked well in capturing the characteristic features of the pressure-temperature phase diagrams of UPd$_2$Si$_2$ and UNi$_2$Si$_2$ \cite{Quirion98}.

Unlike those studies, for the magnetic structure modulation along the $a$-axis, like realized in \uas, in-plane AFM interactions are essential.
Sugiyama {\it et al.} deduced the $(++-)$ structure in an external magnetic field, by treating an Ising model with a mean-field calculation \cite{Sugiyama90}.
For that model, they assumed magnetic moments localized on the uranium site, which frustrate via AFM exchange interactions within third neighbors.
Very recently, Farias {\it et al.} studied a Heisenberg Hamiltonian with the same type of magnetic frustration in a more general and detailed manner, and found that magnetic order with a propagation vector parallel to the $a$-axis can be stabilized even without magnetic field \cite{Farias16}.
This ground state appears when the intersite AFM interaction along the [110] direction is dominant, according to a phase diagram which they calculated.
These studies imply that such kind of magnetic frustration underlies magnetic correlations in URu$_2$Si$_2$ and \uas.

The magnetism in U$T_2$Si$_2$ compounds is considered to be ruled by the RKKY interaction, which modulates in real space with a wave number that depends on the Fermi wave number of the electronic system.
Hence the electronic structure and the interatomic distance are the essential factors to yield the magnetic frustration.
In spite of the potential similarity of the magnetic correlations, the electronic structures should be rather different between URu$_2$Si$_2$ and \uas.
From a number of studies, it is considered that 5f electrons substantially hybridize with conduction electrons in URu$_2$Si$_2$ \cite{Fujimori17, Yanagisawa13, Endstra93, Mydosh_Rev14}.
On the other hand, there is no experimental report on the electronic structure of \uas, but the weaker hybridization effect is expected from a theoretical point of view; the 5d levels of Au located deeply below the Fermi level may reduce the hybridization of 5f electrons with conduction electrons via f-d hybridization \cite{Endstra93}. 
In contrast to the dissimilarity of the electronic structures, it is noticeable that both the Au system and the Ru system have relatively long $a$-axes among the U$T_2$Si$_2$ family.
Experiments with hydrostatic pressure and uniaxial stress on \uas \ may bring interesting information on the magnetic correlations that cause the $(++-)$ magnetic structure.

\section{Conclusions}
Our neutron diffraction study revealed that the AFM order with the propagation vector or $\bm{k} = (2/3, 0, 0)$ takes place in \uas, which was the last system in the U$T_2$Si$_2$ family whose magnetic ordered state had not been unveiled.
We could not detect the small FM component which was observed in the bulk magnetization, because of the experimental accuracy.
Considering the FM component and the AFM component observed respectively in the magnetization and the neutron diffraction, together with the previous NMR results, we concluded that the magnetic structure of \uas \ is the $(++-)$ structure along the $a$-axis with magnetic moments pointing parallel to the $c$-axis.
The structure resembles that realized in Rh-doped and nondoped URu$_2$Si$_2$, suggesting similar magnetic correlations in these systems.
Further studies including measurements of details of the ferromagnetic component and magnetic excitations in \uas \ are in progress.

\begin{acknowledgments}
The present research was supported by JSPS Grants-in-Aid for Scientific Research (KAKENHI) Grant No. 15K05882 and 15K21732 (J-Physics), Strategic Young Researcher Overseas Visits Program for Accelerating Brain Circulation, and the Czech Science Foundation Grant No. P204/16/06422S.
The studied single-crystal sample has been prepared and characterized in the Czech Research Infrastructure Materials Growth and Measurement Laboratory (https://mgml.eu/).
The work was also supported within the program of Large Infrastructures for Research, Experimental Development and Innovation (Project No. LM2015050) and research project LTT17019 financed by the Ministry of Education, Youth and Sports, Czech Republic.
\end{acknowledgments}

\end{document}